# Information-Adaptive Denoising for Iterative PET Reconstruction

André Salomon[1], Andriy Andreyev[2], and Andreas Goedicke[1]

[1]Philips Research, Department of Oncology Solutions, High Tech Campus 34, 5656 AE Eindhoven, The Netherlands
[2]was with Philips Healthcare, Highland Heights, OH, USA, he is now with Carl Zeiss X-Ray Microscopy, Pleasanton, CA, USA

## ABSTRACT

Quantitative accuracy and thus diagnostic precision in Emission Tomography is impaired by the inherent random characteristics of the data acquisition leading to statistical image noise. Edge preserving spatial variation regularized iterative image reconstruction approaches such as using relative difference prior (*RDP*) require case-specific control parameter adaptation for optimized local contrast-vs-noise tradeoff.

For MLEM-type reconstruction, we propose and evaluate *iRDF*, which automatically adapts RDP edge preservation parameters according to local image noise and PET data characteristics. In order to more effectively distinguish between clustered noise spots and small isolated tumors, we introduce a neighborhood-difference-based hot-spot artifact correction based on minimum spatial-information threshold. The proposed method was evaluated using NEMA IQ phantom data as well as clinical patient data.

After initial *iRDF* base parameter tuning, all datasets were reconstructed with the same setup. The results showed that *iRDF* maintained similar image quality regardless of statistics without requiring manual parameter tuning, in contrast to e.g. RDP. With NEMA-IQ phantom data, local image variance was reduced to ~33%, while contrast of small spheres could be mostly preserved compared to non-regularized OSEM. Using a quarter of the originally acquired list-mode data, a noise decreased to ~22% while SUV-max has been reduced to ~75% of OSEM-based results. NEMA phantom and clinical data showed improved signal-recovery-to-noise ratios, leading to an overall ~3 times higher feature detectability especially in small lesions. Finally, the processed examples illustrate the effectiveness of the proposed hot-pixel artefact correction.

We therefore conclude that proposed auto-adaptive *iRDF* regularization demonstrates its high potential to effectively reduce the existing burden of prior parameter tuning. It realizes a reasonable trade-off between feature contrast and image noise on both local and global scale. Moreover, according to the increased noise robustness at different count statistics, *iRDF* can be considered an interesting alternative especially for low-dose PET imaging applications.

## I. INTRODUCTION

As a clinically established diagnostic tool, molecular imaging and in particular Positron Emission Tomography (PET) has been significantly improved during the past decades in terms of spatial resolution, sensitivity, and quantitative accuracy. The image reconstruction process aims to efficiently convert gamma particle detection events registered in a dedicated (ring) detector gantry into an accurate estimate of the underlying radiotracer spatial distribution inside the patient. For gamma detection, each detector element is equipped with a scintillation crystal (e.g., LYSO) and an attached (multi-channel or silicon) photomultiplier. In clinical applications – using specifically selected radiotracers with metabolically or receptor-affine ligands – the resulting visual (2D/3D) representation of the radiotracer distribution is used to analyze in-vivo biological processes linked to specific diseases. In oncology-related diagnostic applications, assessing obvious deviations in the “normal” (i.e. physiological) tracer-accumulation in the image is used to reveal a locally increased glucose uptake, as present in hyper-mitotic, cancerous lesions.

Achievable accuracy substantially depends on the amount of collected information from the underlying PET tracer distribution, i.e. the detected number of coincidently emitted photon-pairs. Compared to CT or X-ray imaging, where relatively high photon count numbers limit the noise impact, noise-induced data variation in emission tomography represents one of the main quality degrading effects, leading to related textural artifacts in the reconstructed images. Within these noise patterns, feature detectability of less dominant small lesions can be significantly reduced, as they sometimes can be hardly distinguished from the present background signal fluctuations.

Noise reduction in iterative reconstruction [1] is most frequently applied using the following methods:

a) The iterative reconstruction is stopped after a pre-defined number of iterations without any individually measured convergence criterion before global image noise reaches an unacceptable level. In PET, this simple, effective and therefore most frequently applied approach leads to visually appealing results, but sub-optimal contrast recovery and spatial resolution in the final PET image (discussed e.g. in [2]). Moreover, it is impossible to find a one-size-fits-all parameter that provides the best image quality for all patient acquisitions and study types.

b) Choosing from a variety of different filter types, the final reconstruction outcome is post-filtered either entirely in spatial domain, frequency domain, or using a combination of both (such as *wavelet*-filtering) [3]. Considering typical reconstruction time requirements, these filters can be applied *"on-the-fly"* which allows adapting and optimizing the outcome according to application or clinician-specific preferences. However, the effectiveness of post-filtering depends on the detail level but also on the already present artifact level of the reconstructed image. Thus, there is a certain risk even using advanced adaptive filtering techniques that artificial features created during the iterative reconstruction process are not only preserved but even further enhanced, or that subtle, real features are suppressed in the final processing step. Moreover, these techniques do not address the inherent problem of inhomogeneous feature convergence and the related (probably impossible) task to stop the reconstruction process at a globally optimal number of iterations.

c) Controlling noise artifact formation is made an integral component of the reconstruction process itself. This can be realized using additional filtering of the intermediate results (e.g. after each full OSEM cycle) as local weighting of the image update, or via intrinsic filtering properties of smooth and overlapping image basis functions (blobs) applied in the 3D radiotracer representation [4].

More advanced methods in the latter category of regularization in PET add a further general constraint $R(\lambda)$ to the objective function, which can be used to steer the reconstruction behavior in various ways. In mathematical terms, PET reconstruction [5][6] can be formally solved via a gradient descent scheme in case of differential convex objective functions:

$$\lambda_j^{n+1} = \lambda_j^n + \eta^n \, d_j(\lambda_j^n) \frac{\partial}{\partial \lambda_j^n} \left( L(\lambda_j^n) - R_j(\lambda_j^n) \right) \tag{1}$$

where $n$ is the iteration index, $\lambda_j$ the estimated activity in voxel $j$, $\eta^n$ the relaxation factor, $d_j(\lambda_j^n)$ a reconstruction state-dependent scaling (normalization) function at image voxel $j$, and $L(\lambda_j^n)$ the log-likelihood of λ given the measured signal $y$.

While there are also methods using prior information and dictionary learning algorithms such as in [7][8][9] and [10], as well as trust optimization transfer functions [11], $R(\boldsymbol{\lambda})$ is most frequently set up to penalize *spatial variation* in reconstructed image intensity between each voxel $\lambda_j$ and its neighbors $\lambda_k$. Using global ($\beta$) and local ($\omega$) weights for the overall effect and the spatial contribution within the considered neighborhood, respectively, the regularization term can further expressed as:

$$R_j(\lambda) = \beta_j \sum_{k \epsilon N_j} \omega_{jk} \sigma(\lambda_j, \lambda_k) \tag{2}$$

This formulation enables optimizing the penalization characteristics according to the desired image properties. For example, choosing a simple *quadratic prior* (i.e. $(\lambda_j, \lambda_k) = (\lambda_j - \lambda_k)^2$ ) leads to strong noise reduction but also a loss in spatial resolution.

A more advanced prior that also includes edge-preservation feature is the *Relative Difference Prior* (*RDP*) [12][13][14][15][16]:

$$\sigma(\lambda_j, \lambda_k) = \frac{(\lambda_j - \lambda_k)^2}{\lambda_j + \lambda_k + \gamma|\lambda_j - \lambda_k|} \tag{3}$$

with parameter $\gamma > 0$ as the Gibbs prior control parameter. According to Nuyts et al. [15], *RDP* applied to the *Maximum Likelihood Expectation Maximization* (*MLEM*) reconstruction can be expressed as:

$$\begin{aligned} \lambda_j^{n+1} = \quad & \left. \lambda_j^n + \frac{\lambda_j^n}{s_j}\frac{\partial}{\partial \lambda_j^n}\left[\sum_{i=1}^{N} y_i \, log\left(\sum_{j=1}^{p} a_{ij}\lambda_j^n\right) - \left(\sum_{j=1}^{p} a_{ij}\lambda_j^n\right)\right] \right\} \boldsymbol{MLEM} \\ & \left. \underbrace{-\frac{\lambda_j^n}{s_j}\frac{\partial}{\partial \lambda_j^n}\left[\sum_{k \in N_j} \beta_{kj}^{*} \frac{(\lambda_j^n - \lambda_k^n)^2}{\lambda_j^n + \lambda_k^n + \gamma|\lambda_j^n - \lambda_k^n|}\right]}_{=\frac{\lambda_j^n}{s_j}\sum_{k \in N_j}\frac{\delta M_{jk}}{\delta \lambda_j^n}} \right\} \boldsymbol{RDP - Penalty\ term} \end{aligned} \tag{4}$$

where $\beta_{kj}^{*} = w_k \cdot \beta_j / \sum_{\forall k} w_k$ is a local penalty-weighting factor, $\frac{\lambda_j^n}{s_j}\sum_{k \in N_j}\frac{\delta M_{jk}}{\delta \lambda_j^n}$ is the *RDP* penalty term, $a_{ij}$ is the system-matrix entry, i.e., the contribution of voxel $j$ to coincidence index $i$, and $s_j$ the estimated total scanner sensitivity at voxel $j$ scaled with the acquisition time $T$ as used in [17]:

$$s_j = T \cdot \sum_{\forall i} \eta_i a_{ij} \tag{5}$$

where $\eta_i$ includes the attenuation along the tube-of-response, crystal efficiency, as well as dead-time and other calibration factors as proposed by Wang et al. in [18]. In case of time-of-flight (TOF) list mode data, $a_{ij}$ does also account for the TOF information of each coincidence event. For simplification, additional correction factors for scatter and random coincidences are not shown in eq. (4). Using the Euclidian distance-motivated approach, $w_k$ is typically chosen as follows:

$$w_k = \begin{cases} 1 & \text{for voxels sharing a face with the center voxel} \\ 1/\sqrt{2} & \text{for voxels sharing only an edge with the center voxel} \\ 1/\sqrt{3} & \text{for voxels sharing only a point with the center voxel} \end{cases} \tag{6}$$

*RDP* penalizes differences between two neighboring image elements $j$ and $k$ if the relative difference $\overline{r_{jk}} = |\lambda_j - \lambda_k|/(\lambda_j + \lambda_k)$ is significantly lower than $1/\gamma$ , where $\gamma$ is a threshold parameter that steers the prioritization between two filter characteristics: quadratic and linear prior strength as shown in [15]. Hence, local activity values $\lambda_j$ are effectively smoothed for

$$\frac{\lambda_j + \lambda_k}{|\lambda_j - \lambda_k|} \gg \gamma \tag{7}$$

and differences between neighboring activity values are preserved for

$$\frac{\lambda_j + \lambda_k}{|\lambda_j - \lambda_k|} \ll \gamma. \tag{8}$$

For values in between, a mixture of quadratic and linear prior is applied, leading to a weighted smoothing behavior.

Accordingly, the resulting image quality significantly depends on the value for parameter $\gamma$ and the applied penalty weighting factors $\beta_j$. For various applications, *RDP* yielded good results, if appropriate values for both parameters $\beta$ and $\gamma$ were set as evaluated e.g. in [13] and [16]. However, since both depend on the workflow/study setup in terms of patient size, injected radiopharmaceutical dose, acquisition time, and also on scanner specific properties (e.g., sensitivity), a two-dimensional parameter optimization has to be performed for each PET study. The optimization is usually done by visual and quantitative inspection of the image quality while "tweaking" the parameters. Since the reconstruction, which takes at least several minutes for each run, needs to be repeated for each modification, this optimization approach is often considered too time consuming for daily clinical routine. The same issue is still present in more advanced algorithms such as in a regularized *BSREM* scheme that reduced the list of parameters to $\beta$ [19].A typical approach for mitigation of the optimization complexity issue is to find a one-fits-all parameter setting which works well for clinical data acquisitions with typical dose and patient size. Another potential option is providing typical settings for each type of patient (e.g., normal, obese, thin), but also here, one will most likely not get the best results out of *RDP*. Other more advanced methods aim at uniform spatial resolution throughout the whole scanner FOV [20] or (e.g.)

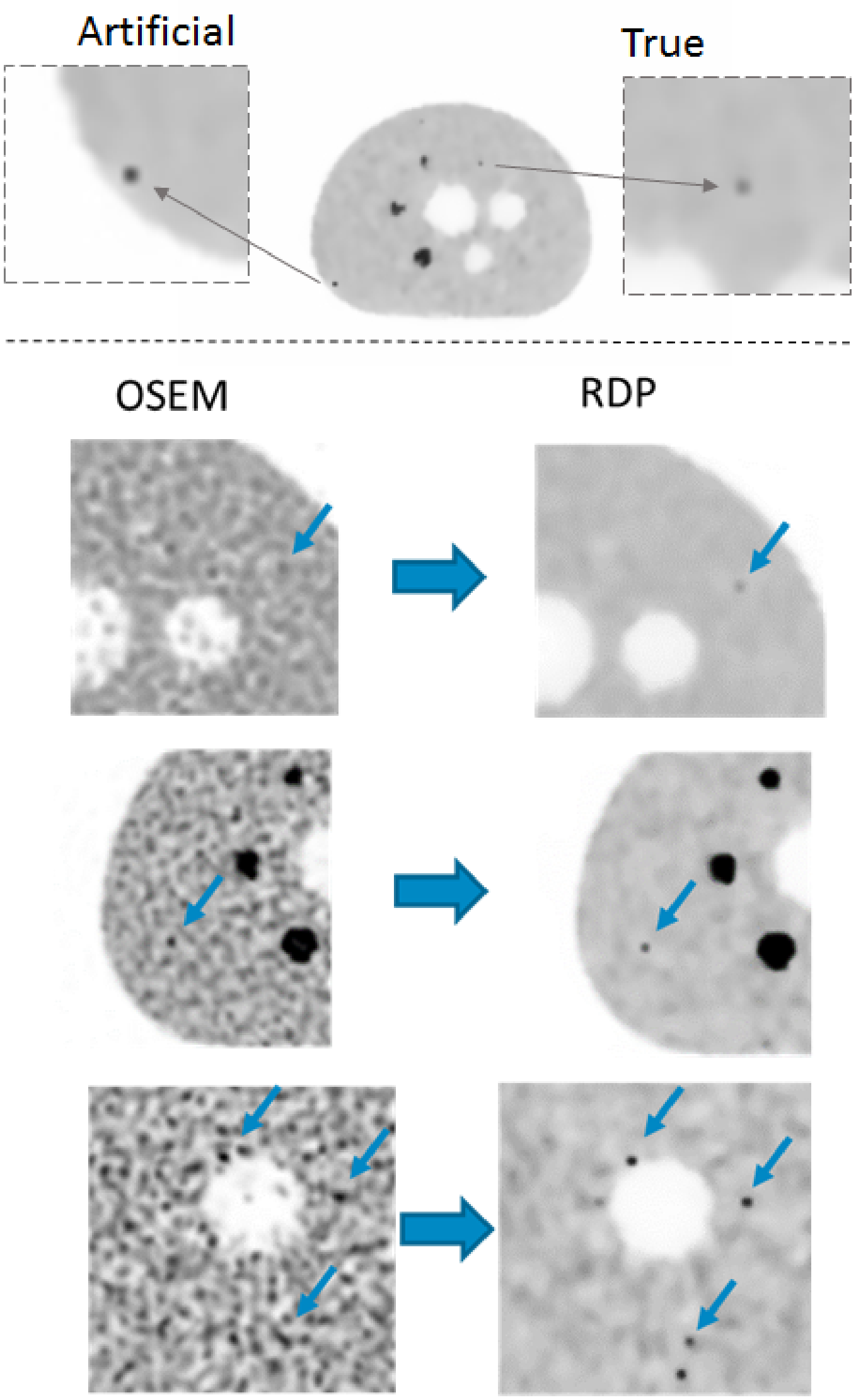

*Fig. 1. First row: Artificial hot lesion ("salt-and-pepper" artifact) in regularized reconstruction (64 iterations of MLEM with RDP) of the NEMA IQ phantom (left) and actual feature (top slice of 10 mm sphere). The images are from the slice centered on the artificial hot-spot with partial view of inserted 10 mm diameter sphere. Bottom images (rows 2,3, and 4) show a comparison of reconstruction results without (left) and with (right) regularization.*

use pre-computed look-up tables of the aggregate certainty map to spatially adopt the smoothing prior in space-variant three-dimensional PET systems [21].

Another general issue with *RDP* is directly related to the selection of $\gamma$, since high values tend to result in artificial hot-spots in the image. These artifacts occur if higher spatial frequencies are not sufficiently penalized by *RDP*. Spatial noise fluctuations randomly exceed the penalty threshold boundary for feature preservation, and additionally are amplified in the resolution recovery section of the reconstruction. Consequently, as most of the noise has been removed from the image by the penalty, these isolated noise pixels/structures may become very prominent in the images during the iteration process and lead to artificial hot lesions as illustrated in Fig. 1. On the other hand, if $\gamma$ is chosen too low, spatial resolution in the image can be significantly decreased.

Since *RDP* aims to combine advanced noise suppression and quantitative accuracy preservation once suitable configuration parameters have been found for each individual PET study, we investigated how *RDP* can be extended in order to realize a dynamic self-adaption of the prior-strength configuration based on the acquired PET data. With this adaptive approach, we strive to completely avoid a time consuming manual parameter optimization for each PET acquisition, leading to a more effective clinical application workflow while avoiding potential human errors in setting reconstruction parameters.

## II. MATERIALS AND METHODS

The *information-adaptive Relative Differences Filter* (*iRDF*) proposed here is based on the fact that image noise formation in *MLEM* depends on the amount of measured samples, i.e., the number of detected decay events. The statistical noise in the image correction factors

$$c_{ij}^n = \frac{\lambda_j^{n+1}}{\lambda_j^n} = 1 + \frac{1}{s_j} \cdot \frac{\partial}{\partial \lambda_j^n} \left[ \sum_{i=1}^{N} y_i \, log \left( \sum_{j=1}^{p} a_{ij} \lambda_j^n \right) - \left( \sum_{j=1}^{p} a_{ij} \lambda_j^n \right) \right] \quad (9)$$

of each MLEM iteration originates from the Poisson noise character of $y_i$ is and its variance is approximately proportional to the local number of detected samples (i.e. $\lambda_j \cdot s_j$) (see also [22][23] and Fig. 2 showing the resulting noise structure after 64 image updates with MLEM without regularization).

We are using the following relationship between statistical noise in the data and the local relative standard deviation $\sigma_{r\,j}$of the statistical noise in $c_{ij}^n$:

$$\sigma_{r\,j} \propto \frac{1}{\sqrt{detected\ decays\ per\ voxel\ j}} \cong \frac{1}{\sqrt{\lambda_j \cdot s_j}} \quad (10)$$

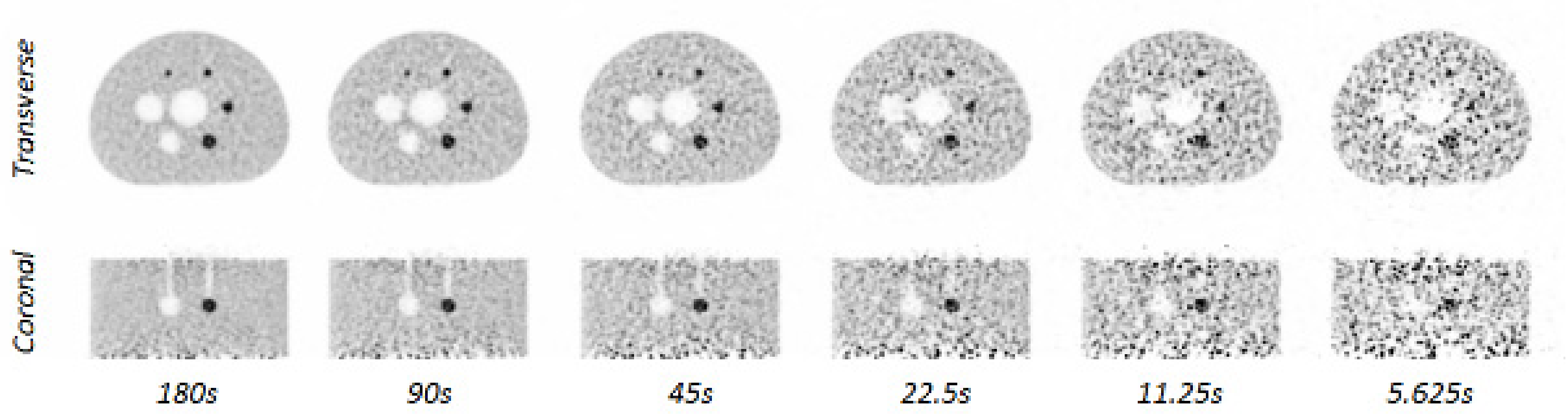

*Fig. 2. Reconstructed activity image using 64 iterations of MLEM of the NEMA IQ phantom for different acquisition times.*

where $s_j$ is defined according to eq. (5). For eq. (10), we assume that the non-local activity barely affects the voxel variance after a certain number of iterations and hence only the local number of detected decays needs to be considered.

In the following, we describe a practical easy-to-use approach towards an automatic and thus very competitive solution as compared to some existing regularization techniques that do not seem to be handy enough for broad clinical use.

### *A.* **Modifications to standard RDP**

The general idea in *iRDF* is to estimate a local value for the $\gamma$ at voxel $j$ (in equation (4)) based on the estimated statistical noise level, i.e., the standard deviation $\sigma_{r\,j}$. Replacing $\gamma$ in (4) with $\gamma^*$ where

$$\gamma_j^* \propto \frac{1}{\sigma_{r\,j}} \tag{11}$$

and considering eq. (10) results in

$$\gamma_j^* = \alpha\sqrt{\lambda_j \cdot s_j} \tag{12}$$

where α is a constant multiplier. As shown in [15] all relative differences $|\lambda_j - \lambda_k|/(\lambda_j + \lambda_k)$ significantly below ${}^{1}/_{\gamma_j^*}$ will be effectively smoothed as if using quadratic prior, while much higher relative differences are treated as real features and preserved in the activity distribution estimated during the iterative reconstruction. The multiplier α is used as a single parameter to steer the noise level in the image. Hence, if α is set to small values, then the image will become very blurry, while big values for α cause image results similar to those without regularization. The ideal value for α in eq. (12) is independent from PET acquisition parameters and individual scanner characteristics since these attributes are already considered by the estimated number of detected decays per voxel $\sqrt{\lambda_j \cdot s_j}$.

Since the estimated local standard deviation is altered with the activity distribution update, $\gamma^*$ is recalculated after each MLEM iteration. Because of this, the original parameter $\gamma$ does not need to be set manually. Instead, it is automatically determined using $\gamma_j^*$ according to the amount of exploitable information contained in the individual PET list-mode data. A somewhat negative side-effect of the proposed modification is that since $\gamma$ now becomes a function of image estimate $\lambda$, formally this breaks the derivation of penalized updating scheme from taking the derivative of the penalty (3). Therefore in the next sections we pay an extra attention to the numerical stability of the proposed algorithm.

For the *RDP* penalty term in (4) the following holds:

$$\frac{\lambda_j^n}{s_j}\frac{\delta M_{jk}}{\delta \lambda_j^n} = -\frac{\lambda_j^n}{s_j}\beta_{kj}^* \frac{(\lambda_j^n - \lambda_k^n)(\gamma|\lambda_j^n - \lambda_k^n| + \lambda_j^n + 3\lambda_k^n)}{(\lambda_j^n + \lambda_k^n + \gamma_j^*|\lambda_j^n - \lambda_k^n|)^2} \tag{13}$$

With the most effective prior shape in terms of smoothing, i.e., with $\gamma_j^* = 0$ and for small differences $\varepsilon \in \mathbb{R}$ with $\lambda_j^n = \lambda_k^n + \varepsilon$, eq. (13) can be written and further simplified using $|\varepsilon| \ll \lambda_j^n$ as follows:

$$
\frac{\lambda_j^n}{s_j}\frac{\delta M_{jk}}{\delta\lambda_j^n} \underbrace{=}_{\gamma_j^*=0} -\frac{\lambda_j^n}{s_j}\beta_{kj}^*\frac{\overbrace{\left(\lambda_j^n-\lambda_k^n\right)}^{=\varepsilon}\overbrace{\left(\lambda_j^n+3\lambda_k^n\right)}^{=4\lambda_j^n-3\varepsilon}}{\underbrace{\left(\lambda_j^n+\lambda_k^n\right)^2}_{=\left(2\lambda_j^n-\varepsilon\right)^2}} \underbrace{\approx}_{|\varepsilon|\ll\lambda_j^n} -\frac{\lambda_j^n\beta_{kj}^*\overbrace{\left(\lambda_j^n-\lambda_k^n\right)}^{=\varepsilon}}{s_j}\cdot\frac{4\lambda_j^n}{\left(2\lambda_j^n\right)^2} \underbrace{=}_{\beta_{kj}^*=w_k\cdot\beta_j/\sum_{\forall k}w_k}
$$

$$
-\frac{1}{s_j}\left(w_k\cdot\beta_j\Big/\sum_{\forall k}w_k\right)\cdot\varepsilon \underbrace{=}_{\beta_j=s_j}\left(w_k\Big/\sum_{\forall k}w_k\right)\cdot\varepsilon \tag{14}
$$

where $\beta_j$ has to be set to $s_j$ in order to perfectly compensate small differences $\lambda_j^n-\lambda_k^n$ in an one-step late application of the prior.

Hence, in order to approach maximum smoothing without producing additional instabilities, especially near the edges of the axial field-of-view (FOV) in PET (where $s_j$ is close to zero), parameter $\beta_j$ is set to $s_j$. This compensates the correction factor $1/s_j$ applied during each update of the activity distribution for commonly used sensitivity correction (see eq. (4): $\lambda_j^n+\frac{\lambda_j^n}{s_j}\frac{\partial}{\partial\lambda_j^n}(\cdots)$) in iterative reconstruction. Choosing otherwise $\beta_j<s_j$, the regularization may become ineffective in compensating the amplified noise added with each (sensitivity corrected) image update, while choosing $\beta_j>s_j$ may over-compensate existing differences between neighboring voxels, leading to unstable convergence behavior.

Including also the modifications for *time-of-flight* (ToF) PET imaging [25] and an one-step-late application of the prior, the resulting update scheme that we used for MLEM with *iRDF* reads as follows:

$$
\begin{aligned}
\lambda_j^{n+½} &= \lambda_j^n+\frac{\lambda_j^n}{s_j}\frac{\partial}{\partial\lambda_j^n}\left[\sum_{i=1}^{N}y_i\,log\left(\sum_{j=1}^{p}a_{ij}\lambda_j^n\right)-\left(\sum_{j=1}^{p}a_{ij}\lambda_j^n\right)\right] \quad \textbf{\textit{MLEM update}}\\
\lambda_j^{n+1} &= \lambda_j^{n+½}-\frac{\lambda_j^{n+½}}{\cancel{s_j}}\frac{\partial}{\partial\lambda_j^{n+½}}\left[\sum_{k\in N_j}\left(\cancel{s_j}\frac{w_k}{\sum_{\forall k}w_k}\right)\frac{\left(\lambda_j^{n+½}-\lambda_k^{n+½}\right)^2}{\lambda_j^{n+½}+\lambda_k^{n+½}+\gamma_j^*\left|\lambda_j^{n+½}-\lambda_k^{n+½}\right|}\right] \quad \textbf{\textit{iRDF smoothing}}
\end{aligned} \tag{15}
$$

where index $n+½$ indicates a sub-step of the whole update scheme $\lambda_j^n\rightarrow\lambda_j^{n+½}\rightarrow\lambda_j^{n+1}$ and

$$
\gamma_j^*=\alpha\sqrt{(G_\sigma(\lambda^{n+½}))_j\cdot s_j} \tag{16}
$$

includes an adapted Gaussian smoothing operator $G_\sigma$ with a local smoothing strength equal to the expected scanner resolution (e.g., 4 mm FWHM for Philips Vereos PET/CT). The reason to introduce $G_\sigma$ is twofold:

1. The spatial accuracy of the estimated activity distribution is limited by the spatial resolution of the scanner which in our case is specified with 4 mm.
2. The smoothing operator supports the convergence of $\gamma_j^*$ and thus in practice prevents a continuous decrease of the local prior strength

The objective function $\Phi(y)$ which has to be optimized can be written as follows:

$$
\begin{aligned}
\Phi(y) &= \sum_{i=1}^{N}y_i\,log\left(\sum_{j=1}^{p}a_{ij}\lambda_j^n\right)-\left(\sum_{j=1}^{p}a_{ij}\lambda_j^n\right)\\
&\quad -\sum_{j=1}^{p}\sum_{k\in N_j}\left(\frac{w_k}{\sum_{\forall k}w_k}\right)\frac{\left(\lambda_j^n-\lambda_k^n\right)^2}{\lambda_j^n+\lambda_k^n+\alpha\sqrt{(G_\sigma(\lambda^n))_j\cdot s_j}\left|\lambda_j^n-\lambda_k^n\right|}
\end{aligned} \tag{17}
$$

The minimum prior strength is applied in case of large absolute differences $\left|\lambda_j^n - \lambda_k^n\right|$ and large $\gamma_j^*$. In this case, i.e., for $\lambda_j^n + \lambda_k^n \ll \gamma_j^* \left|\lambda_j^n - \lambda_k^n\right|$ and without applying a smoothing operator $G_\sigma$ (i.e. $(G_\sigma(\lambda^n))_j \rightarrow \lambda_j^n$) eq. (17) simplifies to

$$\Phi(y) = \sum_{i=1}^{N} y_i \, log\left(\sum_{j=1}^{p} a_{ij}\lambda_j^n\right) - \left(\sum_{j=1}^{p} a_{ij}\lambda_j^n\right) - \sum_{j=1}^{p}\sum_{k\in N_j}\left(\frac{w_k}{\sum_{\forall k} w_k}\right)\frac{\left|\lambda_j^n - \lambda_k^n\right|}{\alpha\sqrt{\lambda_j^n \cdot s_j}} \tag{18}$$

Here, the gradient of our prior is $\frac{d}{d\lambda_j^n}\frac{\left|\lambda_j^n-\lambda_k^n\right|}{\alpha\sqrt{\lambda_j^n \cdot s_j}} = \frac{1}{\alpha\sqrt{s_j}} \cdot \left[\frac{\lambda_j^n-\lambda_k^n}{\sqrt{\lambda_j^n}\left|\lambda_j^n-\lambda_k^n\right|} - \frac{\left|\lambda_j^n-\lambda_k^n\right|}{2\left(\lambda_j^n\right)^{3/2}}\right]$. Further for $\lambda_j^n \gg \lambda_k^n$ (i.e. for a high contrast lesion at $j$) the gradient finally becomes $\frac{1}{\alpha\sqrt{s_j}} \cdot \left[\frac{1}{\sqrt{\lambda_j^n}} - \frac{1}{2\sqrt{\lambda_j^n}}\right] = \frac{1}{2\alpha\sqrt{\lambda_j^n s_j}}$. Hence while increasing the number of detected counts $\lambda_j^n s_j$ to infinite, for large differences the *iRDF* approaches zero and then locally converges to a simple MLEM algorithm without prior. This allows to directly apply *iRDF* to noise-free data where $s_j$ is infinite (i.e. infinite acquisition time and finite scanner sensitivity) and the application of a prior term is superfluous and unnecessarily diminishes the spatial resolution. In contrast to this, *RDP* behaves similar like a linear prior for large differences as shown by Nuyts et al. in [15], i.e., with a constant gradient.

For practical reasons and simplification of the algorithm, $\gamma_j^*$ is calculated "one step late", i.e., it is based on the estimated activity distribution of the previous iteration results similar to e.g. the scatter correction (SSS) which has not been explicitly shown in eq. (15). Hence, for the derivation $\frac{\partial}{\partial\lambda_j^n}$ in eq. (15), the value for $\gamma_j^*$ is not a function of $\lambda_j^n$ but set to a constant as used in eq. (13). Therefore, strictly speaking we can no longer call *iRDF* method a penalized reconstruction from classical theory of regularized reconstruction point of view.

### *B.* Adaptation of RDP to Blob-based reconstruction

The relation between neighbored image elements can be more naturally described using a non-voxel spatial representation as discussed in [26] and [27]. Thus, we integrated it into the Philips PET image reconstruction code [18] including iterative list-mode reconstruction with full use of the time-of-flight information [28] (e.g., ~590 ps resolution for older PET/CT systems and ~320 ps for recent digital PET/CT systems). The reconstructed tracer activity distribution is represented using a regular grid of overlapping Kaiser-Bessel basis-functions (also known as *blobs*). The blob grid is configured according to physical/design properties (such as the scintillation crystal pitch in the detector modules) to optimally support the tradeoff between spatial resolution and image noise. In the current implementation, blob-centers are located in a body-centered cubic (BCC) grid [6], generated by a combination of two regular (cubic) grids (one shifted relative to the other by half grid pitch in all 3 dimensions) instead of a denser but also computationally more expensive hexagonal grid structure. For the BCC grid arrangement in our test implementation, we set a 4.4 mm uni-directional center pitch for each cubic grid and a blob radius of 6 mm. Further *RDP* modifications were required regarding the prior weighting scheme for incorporating neighboring volume elements' intensities. According to the geometrical BCC setup, the Euclidian distance-motivated weighting scheme as described in eq. (6) was adapted for use in eq. (15) with a blob-based neighboring scheme as follows:

$$w_k = \begin{cases} 1 & \text{if } k \text{ is in the same grid as } j \\ \sqrt{3}/2 & \text{if } k \text{ is in the neighboring grid to } j \end{cases} \tag{19}$$

where $j$ is now representing the index in the blob-domain. Index $k$ would run over 6 closest blobs from the same grid as $j$, and over 8 closest blobs from the neighboring grid as shown in Fig. 3.

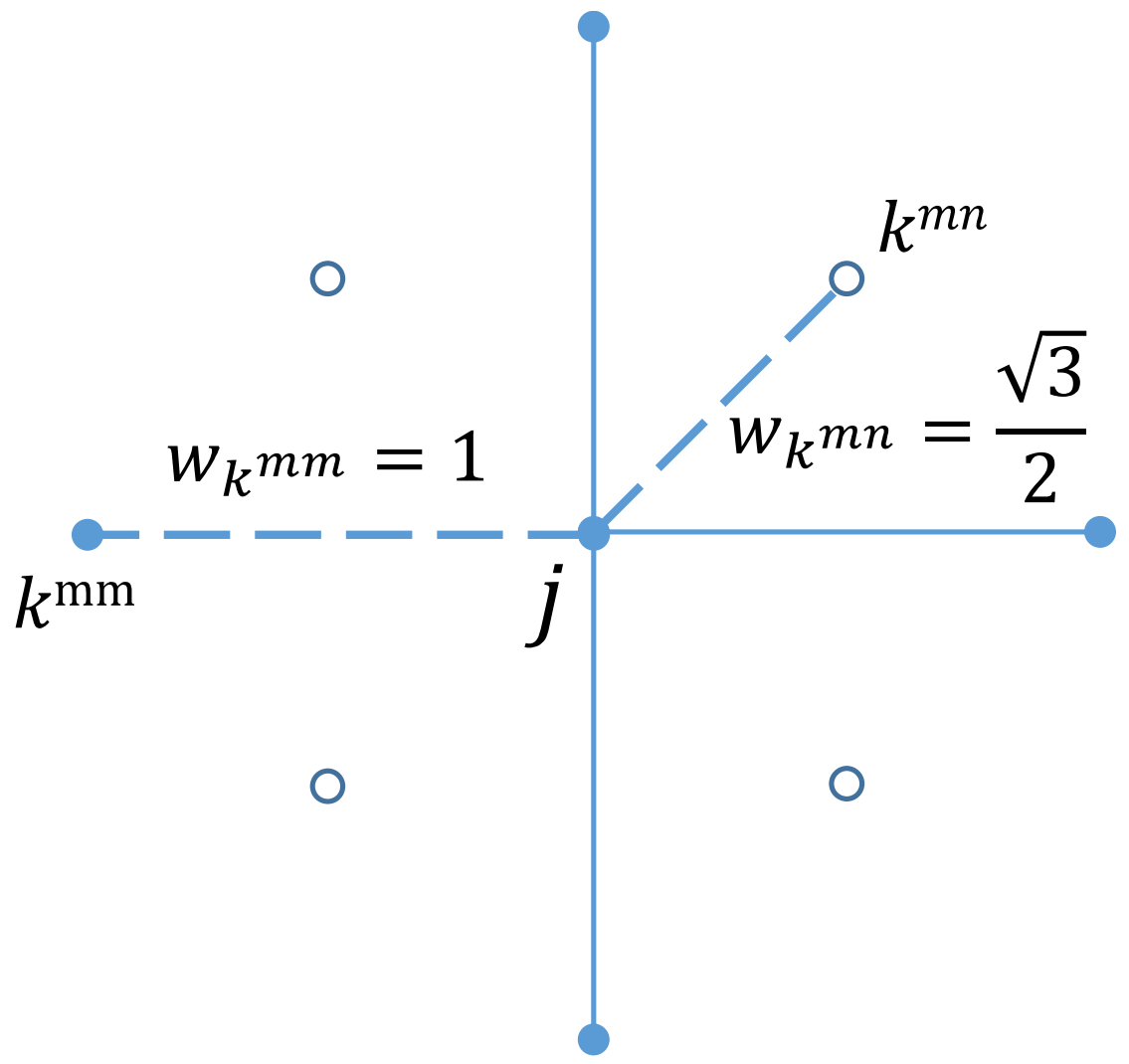

*Fig. 3. Schematic two-dimensional illustration of the weighting scheme for the BCC grid. Filled and open circles represent blob volume elements from two interlaced simple cubic grids in 3D.*

### *C.* One-time Calibration of iRDF

The new parameter $\alpha$ which has been introduced in IIA can be used to steer the general image quality and should be set to a value which suppresses noise as much as possible while preserving the contrast in small lesions. In the following, $\alpha$ is optimized to a one-size-fits-all value that should approximately provide the best detectability of lesions with a pre-defined size (e.g., ~10 mm diameter) for each individual acquisition. In order to do so, the former NEMA-type phantom study was used, and the contrast-recovery-to-noise-ratio (*CRNR*) was defined by:

$$CRNR_{hot} = \underbrace{\frac{\bar{\lambda}_{sphere} / \bar{\lambda}_{background} - 1}{c_{ref} - 1}}_{contrast\ recovery\ coef} \cdot 100\% \cdot \frac{1}{\sigma_{background}} \tag{20}$$

for hot spheres using the reference contrast value $c_{ref} = 4$ and

$$CRNR_{cold} = \left(1 - \bar{\lambda}_{sphere} / \bar{\lambda}_{background}\right) \cdot 100\% \cdot \frac{1}{\sigma_{background}} \tag{21}$$

for cold spheres have been evaluated as an indicator for lesion detectability, with $\bar{\lambda}_{sphere}$ and $\bar{\lambda}_{background}$ representing the average reconstructed intensity values in the spherical ROIs and in the background, respectively, and $\sigma_{background}$ the standard deviation in the background ROI. Deviating from the original NEMA protocol [24], 3D volumetric spherical and cylindrical ROIs have been defined according to Fig. 4. The diameters of the spherical ROIs were set according to the inner diameter specifications for the active spheres of the NEMA NU-2 image quality phantom [24] (i.e. 10, 13, 17, 22, 28, and 37 mm).

In order to find a reasonable setting for parameter $\alpha$ in the *iRDF* formula (16), the phantom data were reconstructed using a variety of different settings between $\alpha = 1$ and $\alpha = 6$, and the corresponding *CRNR*, as well as the NEMA IQ contrast recovery coefficients (CRC) for each sphere (see [24]) were calculated. The latter is important for quantitative analyses such as the assessment of the standardized uptake value (SUV) typically evaluated in clinical applications, such as in initial tumor staging and treatment follow-up studies.

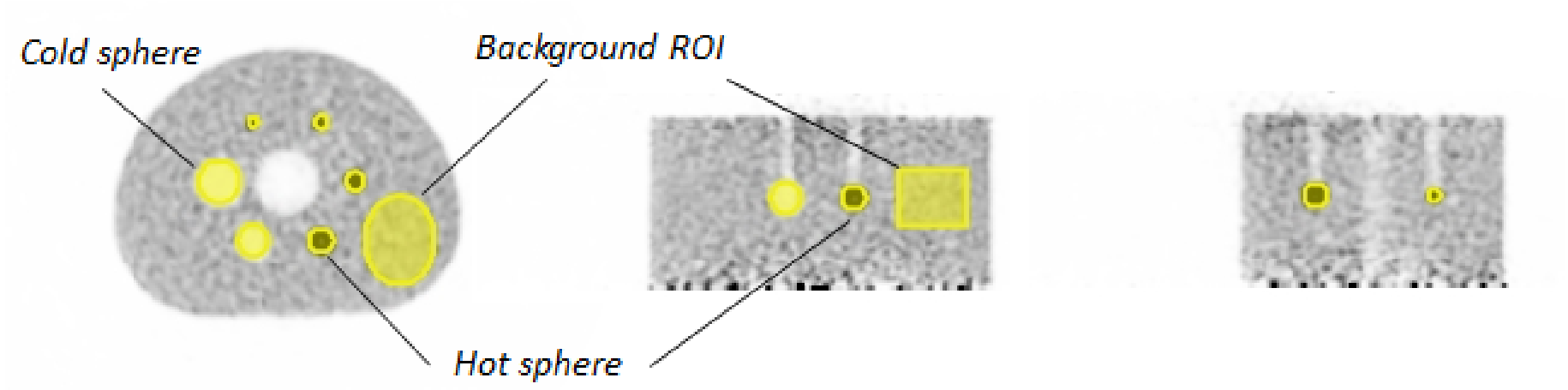

*Fig. 4. Transverse (left), coronal, and sagittal (right) section of the reconstructed activity image of the NEMA IQ phantom (64 iterations of MLEM). Yellow areas indicate Regions-of-Interest (ROIs) for signal-to-noise analysis. 6 spherical ROIs correspond to each of the spheres, while another ROI (elliptic cylinder) represents the background noise.*

### *D.* Detection and removal of artificial hot-spots

As mentioned before, a general problem of *RDP* is reasonably distinguishing between noise and actual features in the image. More conservative parameter settings (e.g. for $\gamma$ and $\beta$ in *RDP*), lead to a reasonably stable algorithmic behavior, but attempting to maximize spatial resolution (and quantitative accuracy of small features) in combination with high noise suppression in the images tends to create hot-spot noise artifacts. These can also be effectively prevented by increasing the smoothing prior strength at small features with high contrast (using lower γ settings), but then resulting SUV values and contrast recovery coefficients will be reduced.

Here, with the automatic parametrization in *iRDF*, we aim to preserve as much spatial resolution as possible while achieving good noise suppression in background regions without features. Therefore, the one-time-calibration described before leads to maximized CRNR for small lesion with ~10 mm diameter, but also introduces hot-sport artifacts (as later illustrated in Fig. 6 and more detailed in Fig. 7 for 45s acquisition time). Having a closer look at those hot-spots, almost all of them are represented by only a single blob in blob basis function space (before the conversion to cubic voxels to obtain the final clinically viewable image takes place). Assuming that any real, trustworthy feature in the image is represented by a certain minimum number of blobs due to finite spatial resolution provided by the imaging device enables a simple approach to identify artificial hot-spots. According to the aforementioned *blob*-grid center arrangement, the minimum diameter of a sphere covering at least two blob-centers is 3.81 mm ($= 4.4\, mm \cdot \sqrt{3}/2$), which is slightly smaller than the scanner's spatial resolution of 4.0 mm in the center of the FOV.

Analyzing initial reconstruction results, we concluded that single hot-blobs exceed their surrounding neighbors by at least ~20%. Thus, using the highest neighborhood blob value (here referred to as $\lambda_{max}$) as reference, blobs exceeding this threshold were capped to $\Delta_{max} \cdot \lambda_{max}$ in order limit the contrast, which is what we call *first-max hot-spot correction*. Formally, this leads to an additional noise mitigation scheme applied after each image update in the form of in-between iterations filtering:

$$\lambda_j \rightarrow \begin{cases} \Delta_{max} \cdot max(\lambda_k|\forall k) & \text{if } \lambda_j > \Delta_{max} \cdot max(\lambda_k|\forall k) \\ \lambda_j & otherwise \end{cases} \tag{22}$$

where $\Delta_{max}$ is set to 1.2 and index $k$ only includes elements neighboring element $j$. Alternatively, in order to remove more spatially extended clustered hot-spots (covering more than one single blob), we also considered a similar comparison between each blob and its second highest neighbor, which is what we further refer to as *second-max hot-spot correction*. Both approaches have been tested for various acquisition times/count levels in the phantom data study.

## *E.* Minimum number of detected decays per image element needed for RDP

In those regions of the reconstructed image where the scanner sensitivity and/or activity concentration are too low, the local relative differences (noise) in the images become too high and are thus preserved by *RDP*. This leads to high local image noise and insufficient smoothing which degrades image quality, especially in those regions with very low sensitivity such as those close to the axial edges of the FOV. For mitigation, we replace (16) in our implementation by the following term:

$$\gamma_j^* = \begin{cases} \alpha\sqrt{(G_\sigma(\lambda))_j \cdot s_j} & if\ (G_\sigma(\lambda))_j \cdot s_j > n_{min} \\ 0 & otherwise \end{cases}, \tag{23}$$

which basically uses a quadratic prior ($\alpha=0$) instead of *RDP* when the estimated number of detected counts, $i.e.,\ (G_\sigma(\lambda))_j \cdot s_j$ is below an effective minimum information threshold value $n_{min}$, where the local statistical information becomes insufficient.

Since the scanner sensitivity and thus the number of detected decays per image element (estimated by $(G_\sigma(\lambda))_j \cdot s_j$) linearly decreases with axial distance from the iso-center, a value for $n_{min}$ has been identified using the NEMA IQ phantom by first analyzing the number of average counts detected at axial image positions where first noise artifacts could be discovered.

## *F.* Evaluation using NEMA IQ phantom and clinical data

For detailed performance evaluation of *iRDF* with acquired phantom data, we used the setup as in the previous sections, i.e., 72 MBq of F-18 FDG, and a total acquisition time of 180s with contrast ratios of 0:4:1 on a Philips Vereos PET/CT scanner. The acquired list-mode data set was then truncated to emulate scans with 90s, 45s, 22.25s, 11.25s, and 5.625s duration. For each case, a reconstruction with and without *iRDF* using $n_{min}$=0 and $n_{min}$=20 have been performed. Moreover, we directly compared non-regularized OSEM (5 iterations/17 subsets) and MLEM (500 iterations) with *iRDF* and standard RDP [15], where the $\gamma$ parameter was set to either enhance the contrast recovery ($\gamma$=100) or to reduce noise in homogenous regions ($\gamma$=20). In order to account for any applied corrections regarding sensitivity, attenuation, and acquisition length, $\beta_j = s_j$ was also used in *RDP*.

For our tests of *iRDF* with clinical data, we reconstructed a clinical whole body PET/CT data set (505 MBq of F-18 FDG, 9 bed positions, each with 90s acquisition time) acquired on a pre-production investigational Philips Vereos PET/CT system. As in the former NEMA investigation, *iRDF* and the original *RDP* are compared to standard non-regularized OSEM.

## III. RESULTS

### *A.* One-time iRDF calibration

The curves shown in Fig. 5 exhibit a noticeable dependency of the contrast and the detectability on *iRDF* parameter α. While the contrast monotonically increases with increasing α, the detectability exhibits a global maximum for each of the acquisition times (rows in Fig. 5) which is between α=2 and α=3.

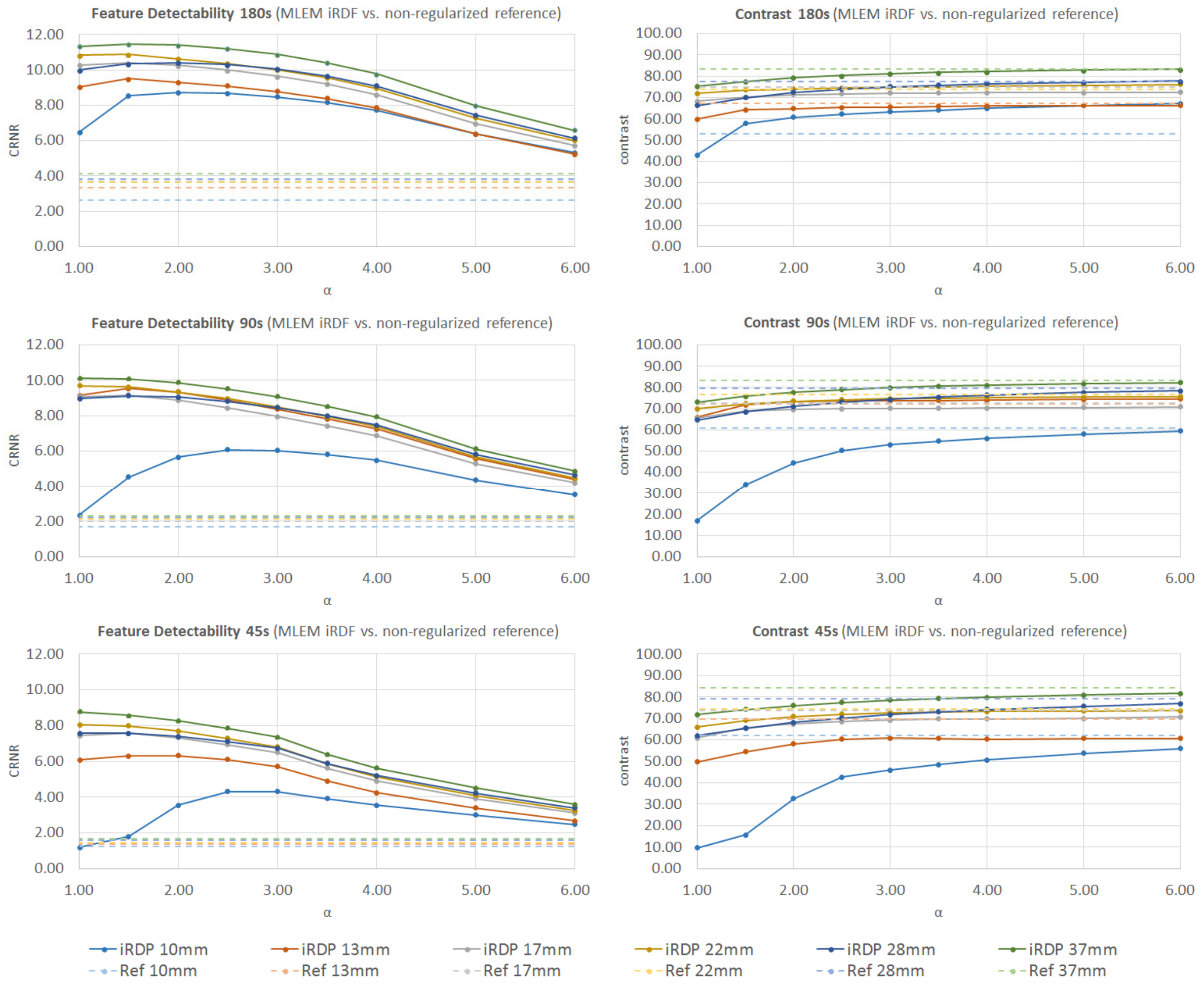

*Fig. 5. α-dependency of feature detectability (left column) and contrast recovery coefficients (right column) using iRDF (solid curves) and non-regularized OSEM (dashed lines) for different sphere diameters. Each row represents a different acquisition time: top: 180s, middle: 90s, bottom: 45s.*

The corresponding reconstructed images are shown in Fig. 6 for the variation of parameter α (right columns) and the reference non-regularized OSEM-based image (first column).

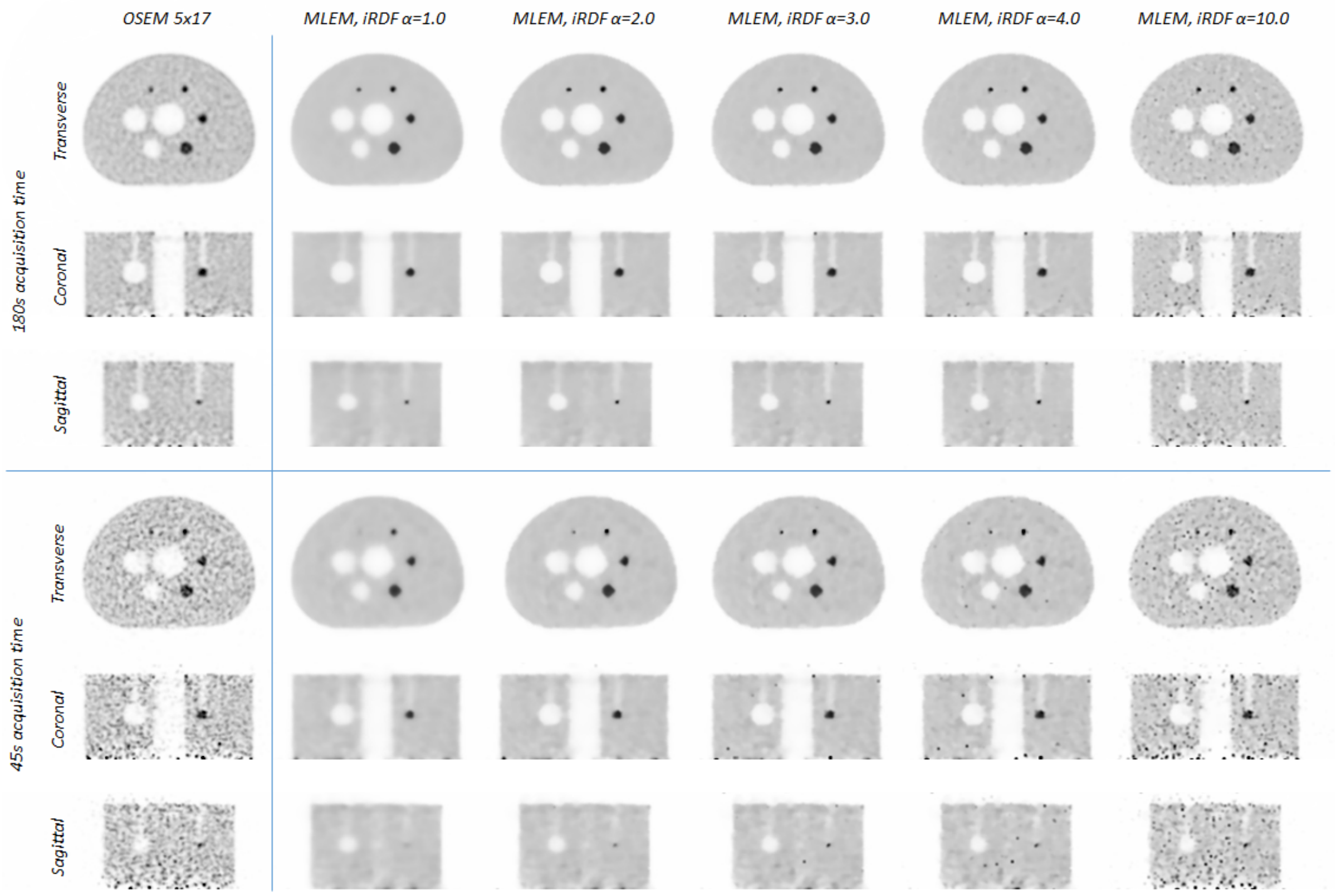

*Fig. 6. Left column shows slices of the reference non-regularized reconstructed image of the IQ NEMA phantom measurement (rows: transverse, coronal and sagittal view for 180s and 45s acquisition time) using OSEM with 17 subsets and 5 iterations, while columns 2 to 6 show the reconstruction results using regularized MLEM and various values for iRDF parameter α.*

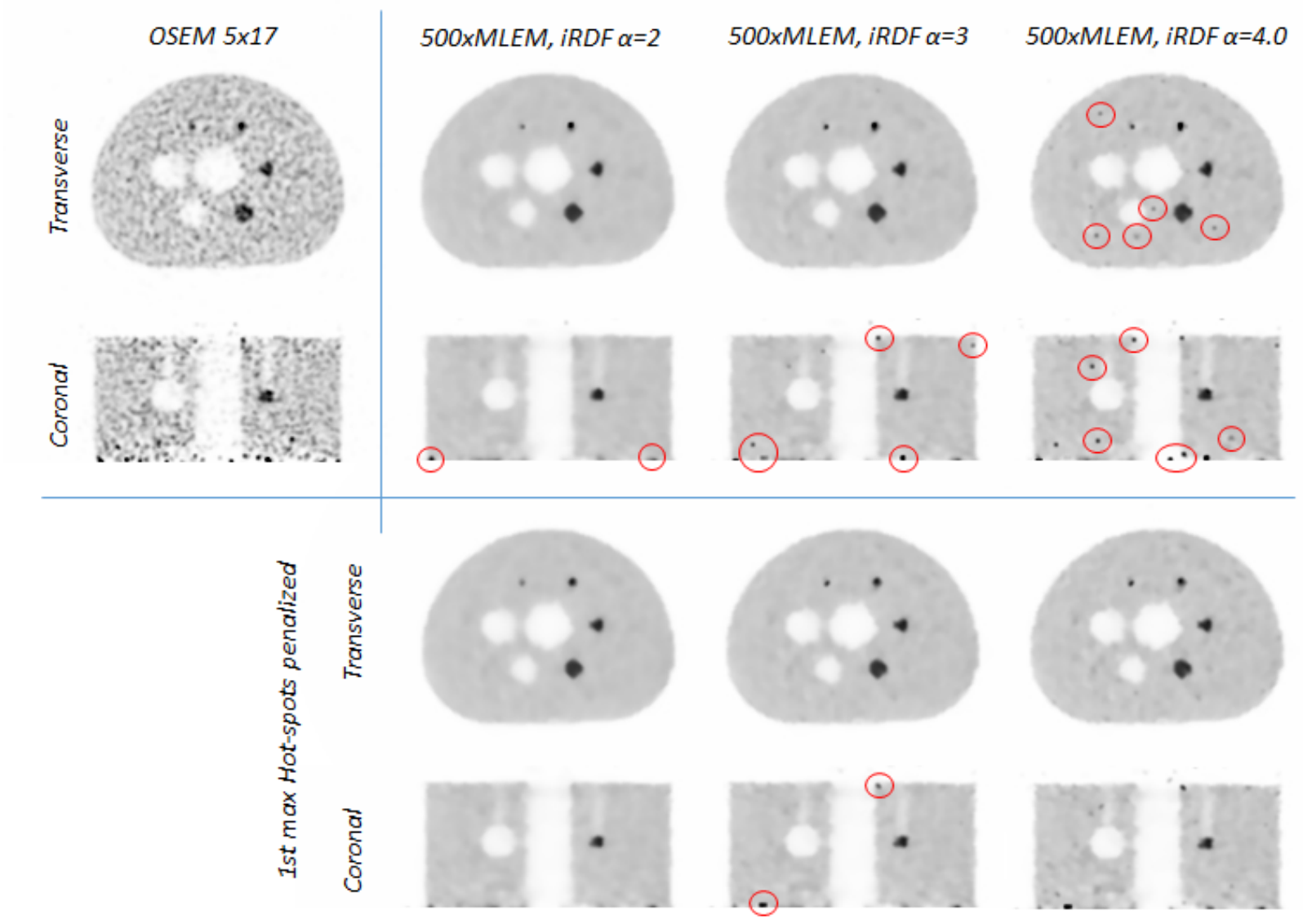

*Fig. 7. Reconstruction results (45s acquisition time) showing introduced hot-spot artifacts with increasing values for α in first two rows, and corresponding results with hot-spot correction in 3rd and 4th row. Artificial hot-spots are highlighted.*

## *B.* Detection and correction of artificial hot-spots

*iRDF* results including the proposed first-max hot-spot penalty are given in the bottom rows of Fig. 7 while the first row shows results without hot-spot correction together with the OSEM-based reference. Additional hot-spot artifacts introduced by step-wise increase of α are highlighted by small red circles.

Fig. 8 shows resulting images of the NEMA phantom using two different types of hot-spot correction with a fixed value for α together with the standard OSEM results for 45s acquisition time.

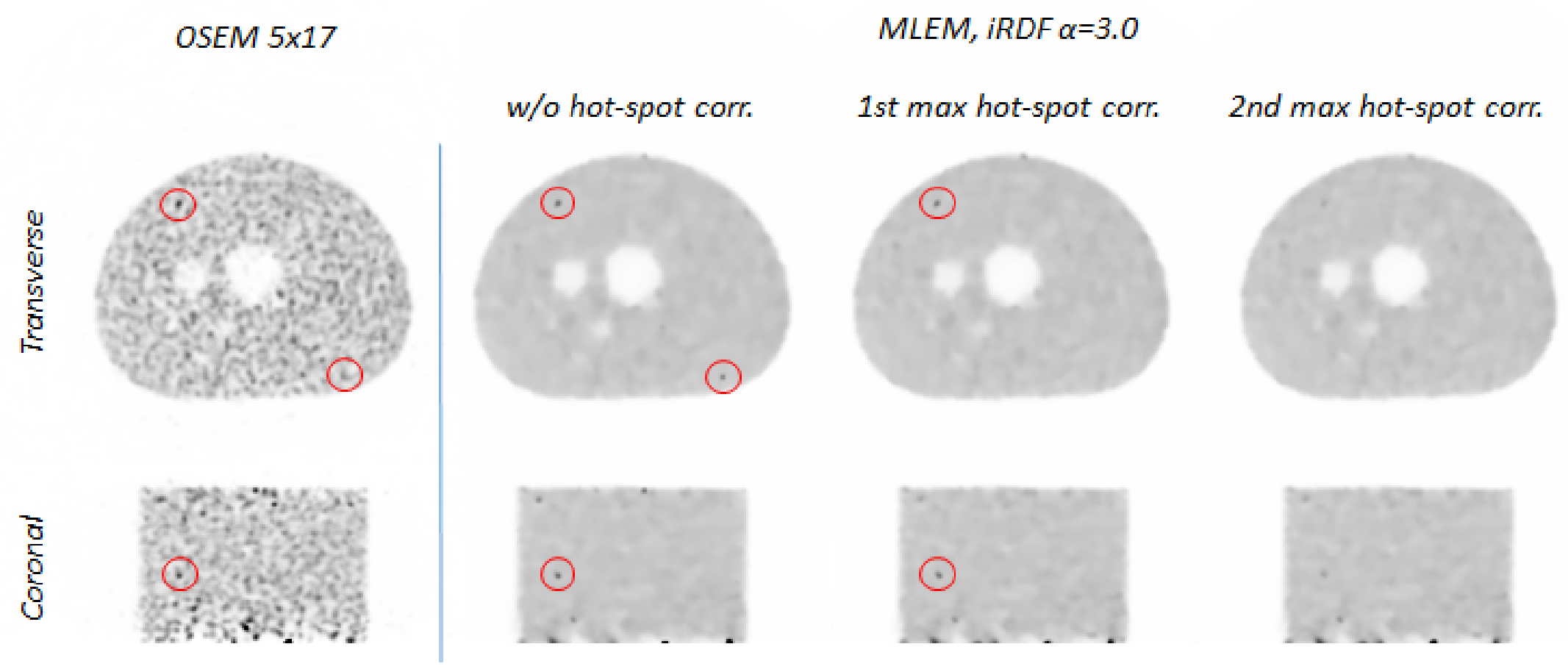

*Fig. 8. Results of the 45s acquisition with iRDF without (left) and with first-max hot-spot correction (center) and second-max hot-spot correction.*

## *C.* Minimum number of detected decays per image element needed for RDP

The second row in Fig. 10 shows the series of reconstructed images of the NEMA IQ phantom that have been used to find the threshold value (and corresponding estimated number of detected coincidences) along the z-axis where the prior did fail to sufficiently smooth the local image. The resulting minimum threshold was found to be 20 counts per image element in $(G_\sigma(\lambda))_j \cdot s_j$.

## *D.* Evaluation using NEMA IQ phantom and clinical data

Following one-time calibration of fixed *iRDF*'s parameters, namely $\gamma$ and the *minimum information threshold* (minimum number of counts per blob), a comparison between standard OSEM reconstruction (5 iterations/17 subsets) and *iRDF* (500 iterations of MLEM) regarding image quality improvement was carried out using both measured phantom and clinical patient data. For various acquisition times, Fig. 10 illustrates the image quality resulting from different regularization configurations on NEMA phantom data.

Similarly, now focusing on 180s and 45s acquisition data, Fig. 9 shows resulting images achieved with non-regularized OSEM, MLEM, standard *RDP* (with γ optimized either for contrast recovery or for noise mitigation), and *iRDF*. Corresponding $SUV_{max}$ and relative standard deviation measures are summarized in Table I and Table II. Considering the same measures, Fig. 11 illustrates the *iRDF* convergence behavior for each defined hot sphere and background ROI (Fig. 4).

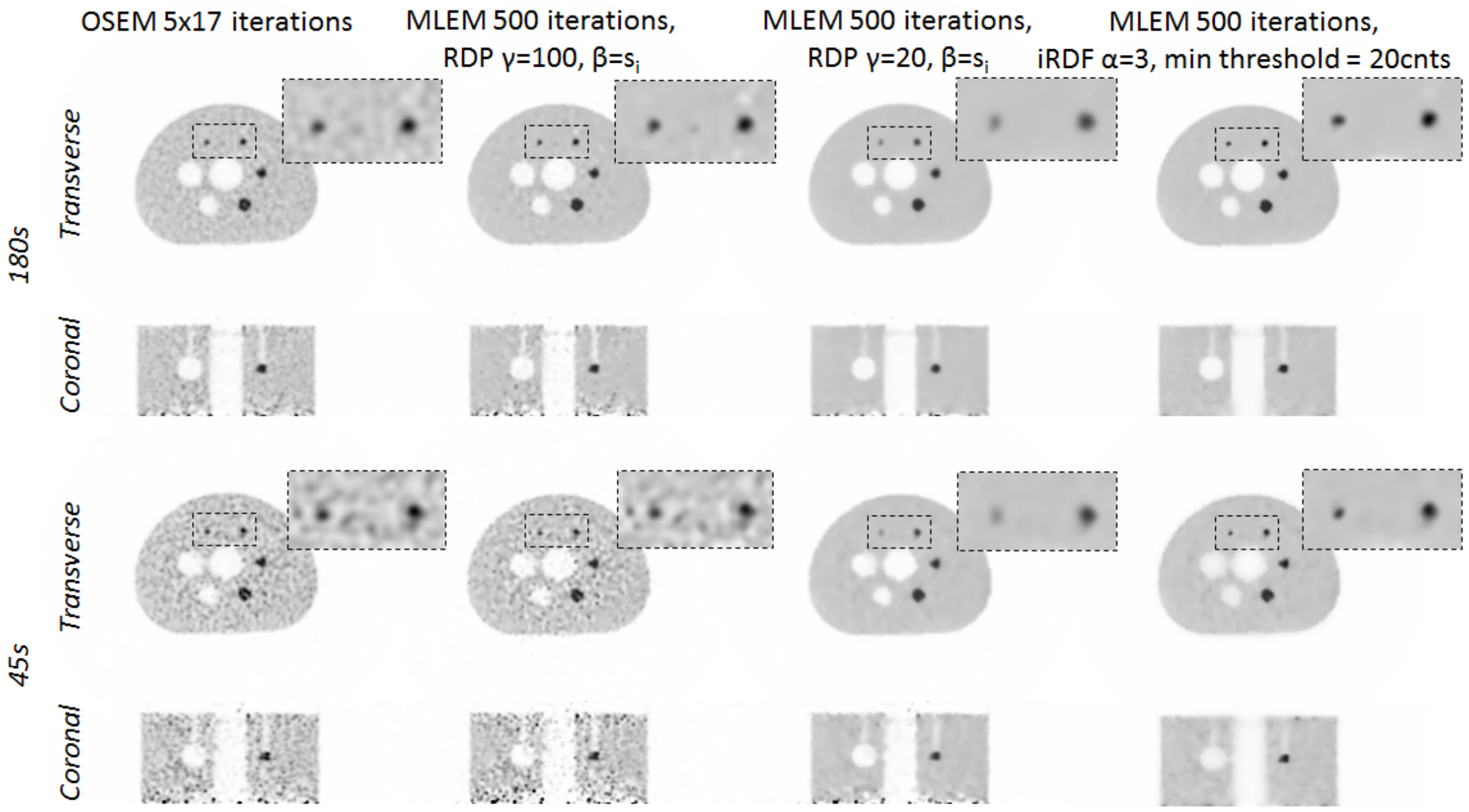

*Fig. 9. Reconstruction results for $4.5{\cdot}10^7$ counts (180s acquisition time, top rows) and $1.125{\cdot}10^6$ counts (45s, bottom rows) for reference reconstruction parameters without regularization (first column), regularized with RDP and high γ (second column), RDP with low γ, and with calibrated iRDF.*

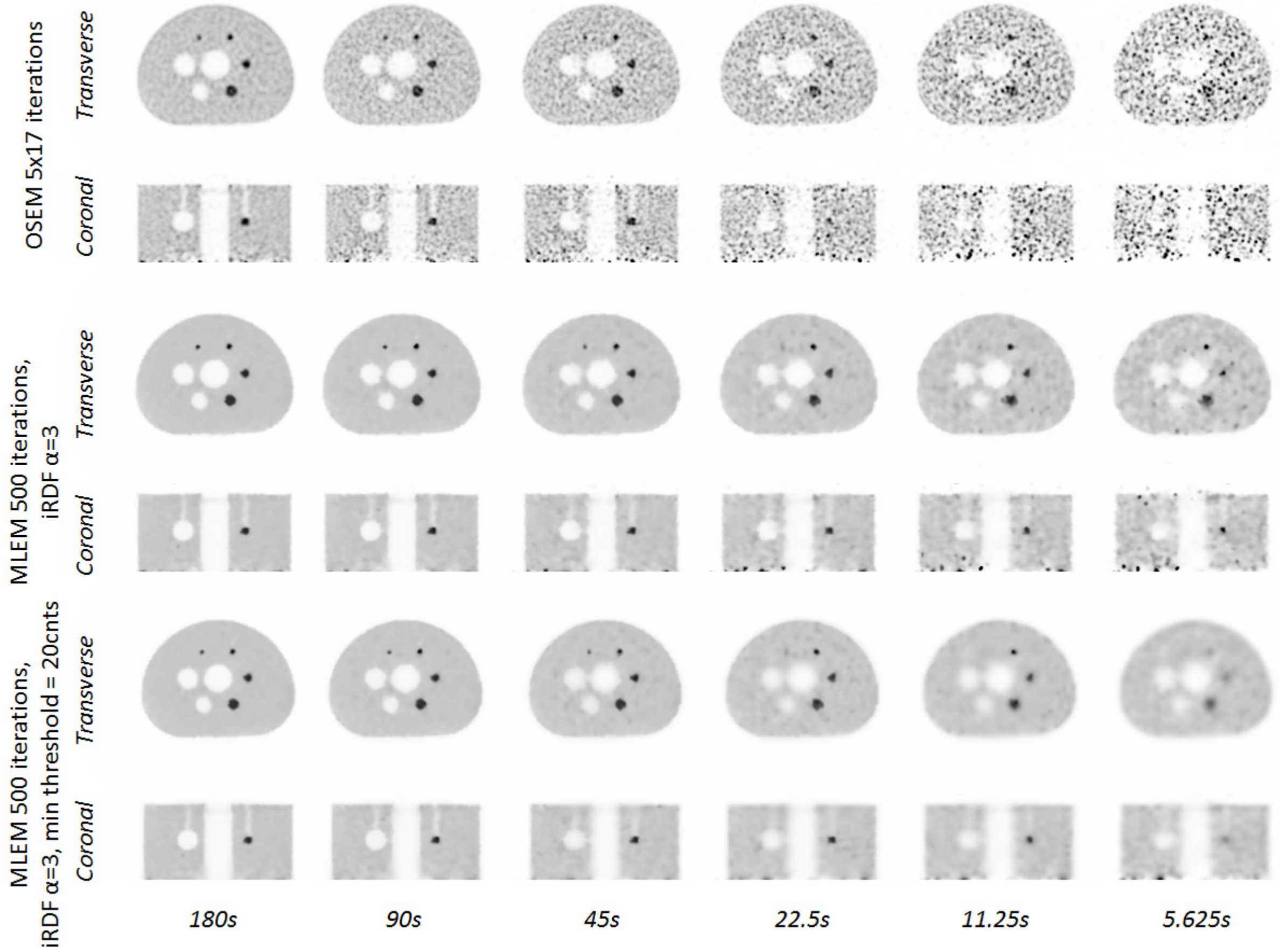

*Fig. 10. Reconstruction results ranging between $4.5{\cdot}10^7$ counts (180s acquisition time) and $1.41{\cdot}10^6$ counts (5.625s) for reference reconstruction parameters without regularization (first row) and regularized with iRDF and 500 iterations of MLEM without information threshold (center) and with minimum information threshold (lower rows).*

TABLE I
QUANTITATIVE VALUES IN RECONSTRUCTED IQ PHANTOM IMAGES (180S)

| | OSEM 5x17 | MLEM 500 | MLEM 500, *RDP* γ=20 | MLEM 500, *RDP* γ=100 | MLEM 500, *iRDF* |
|---|---|---|---|---|---|
| $SUV_{max}$ in 10 mm sphere | 4.0 | 4.9 | 2.6 | 4.3 | 4.3 |
| $SUV_{max}$ in 13 mm sphere | 5.0 | 5.1 | 3.6 | 4.8 | 5.0 |
| Rel. standard dev. in background | 16.1% | 35.2% | 4.5% | 8.5% | 4.9% |

TABLE II
QUANTITATIVE VALUES IN RECONSTRUCTED IQ PHANTOM IMAGES (45S)

| | OSEM 5x17 | MLEM 500 | MLEM 500, *RDP* γ=20 | MLEM 500, *RDP* γ=100 | MLEM 500, *iRDF* |
|---|---|---|---|---|---|
| $SUV_{max}$ in 10 mm sphere | 4.2 | 3.8 | 2.0 | 3.9 | 3.1 |
| $SUV_{max}$ in 13 mm sphere | 6.2 | 6.0 | 3.8 | 6.2 | 5.2 |
| Rel. standard dev. in background | 30.9% | 59.2% | 6.8% | 29.0% | 7.0% |

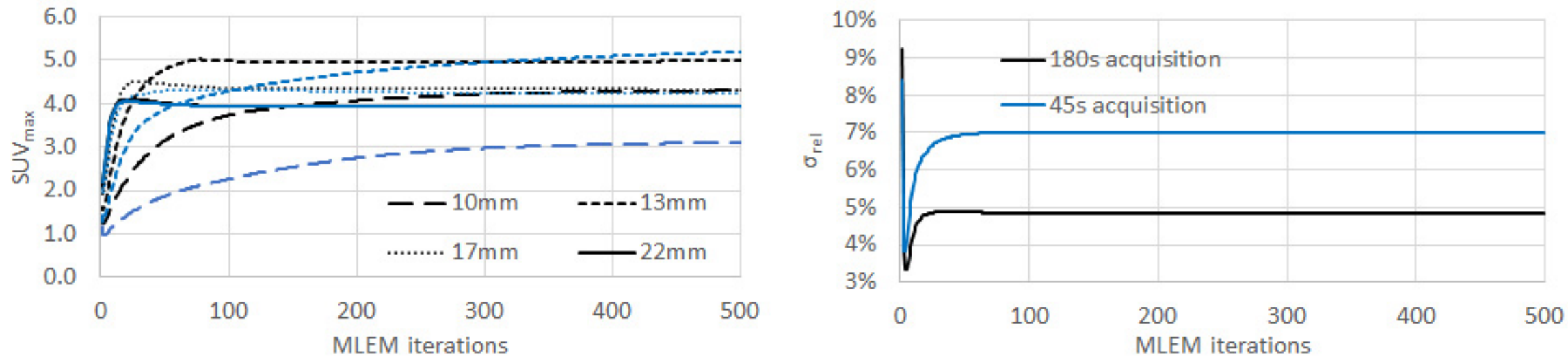

*Fig. 11. iRDF convergence curves of the $SUV_{max}$ value for each hot sphere (left) and convergence of the relative standard deviation in the background region (right) for 180s acquisition (black) and 45s acquisition (blue).*

With the maximum radioactivity concentration $C_{max}$ in the tumor region (measured in Bq/ml), the $SUV_{max}$ was calculated according to

$$SUV_{max} = \frac{C_{max}}{\overline{D}/M} \tag{24}$$

where $\overline{D}$ is the average dose during the PET acquisition (in Bq), and $M$ the mass (in grams). Further examples comparing *iRDF* with both original *RDP* and non-regularized OSEM reconstruction of patient data are presented in Fig. 12 for full acquisition time (now 90s per bed position) as well as for a quarter of the available coincidence data (i.e. simulating a 22.5s acquisition time per bed position). For *RDP* parameter γ, two different values have been used in order to get optimized results with respect to either tumor contrast (*) or noise reduction (**):

a) γ=155 which is equal to average *iRDF's* $\gamma_j^*$ in tumor regions (*)

b) γ=67 which is equal to average *iRDF's* $\gamma_j^*$ in the liver region (**)

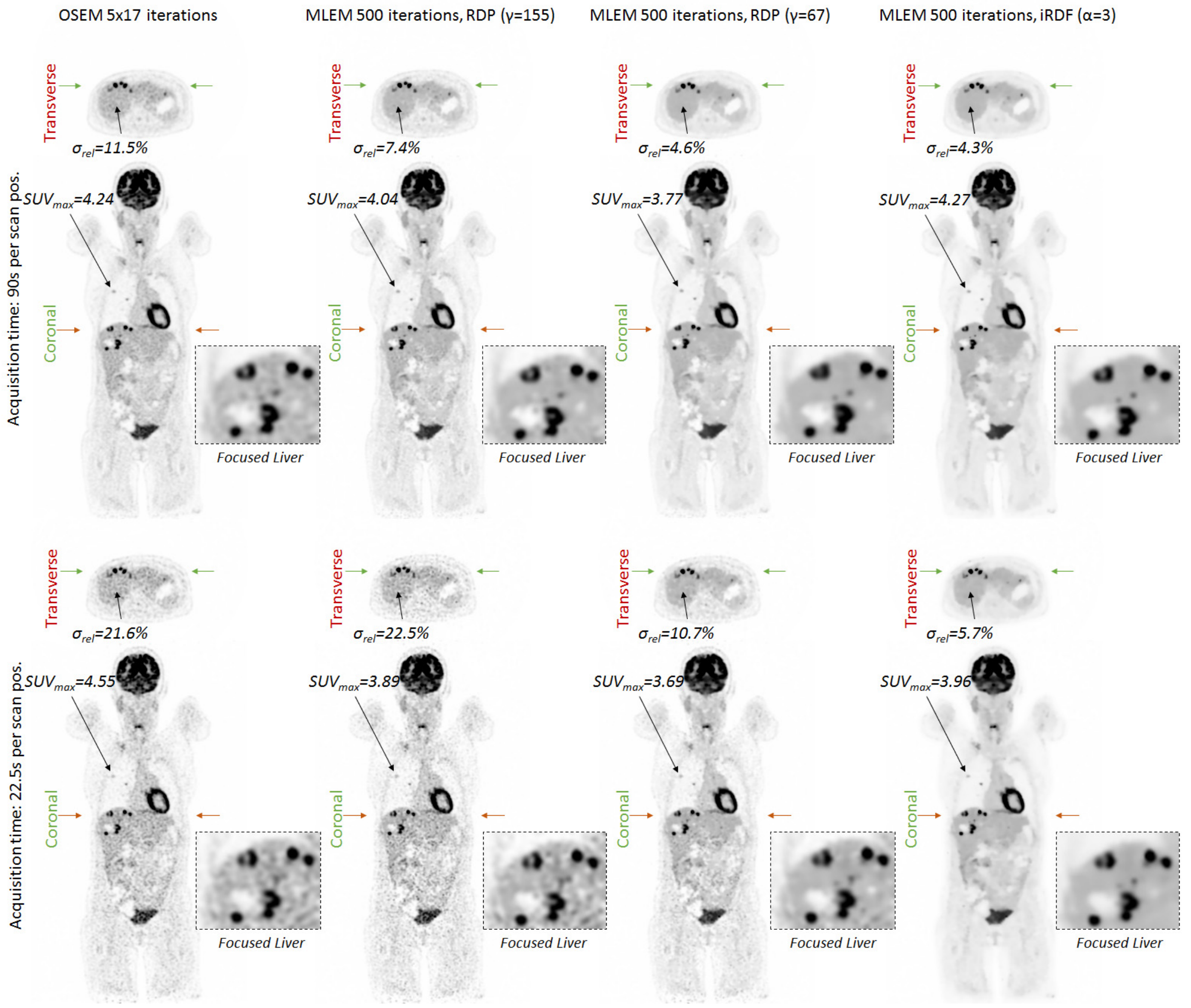

*Fig. 12. Reconstruction results of clinical data with reference reconstruction (first column), regularized with two different settings of RDP (second and third column), and with iRDF (most right column) using the whole data set (top) and a quarter of the available count statistics (bottom).*

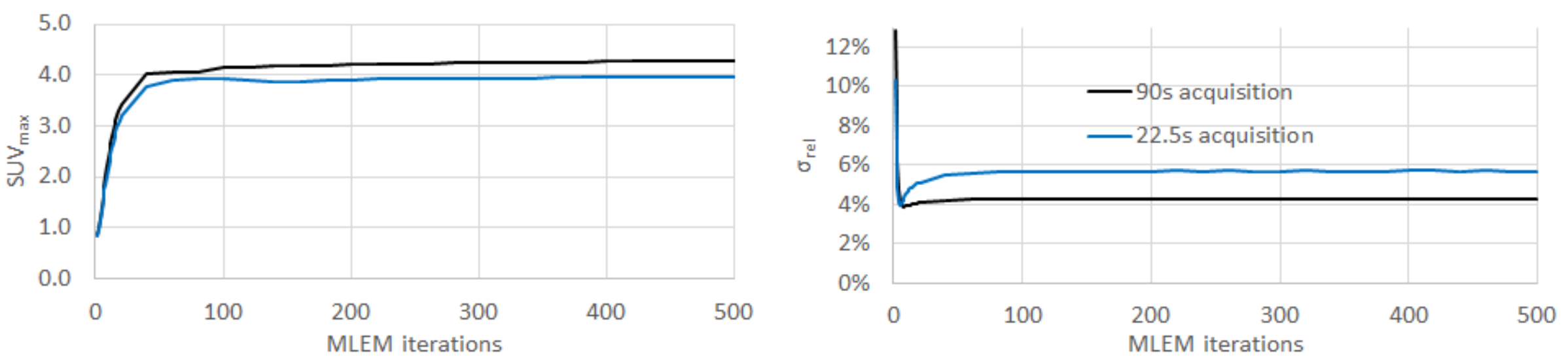

Fig. 13. iRDF convergence curves of the $SUV_{max}$ value for the tumor (left) and corresponding convergence of the relative standard deviation in the liver (excluding tumors) for 90s and 22.5s acquisition time per scan position, i.e., full and quarter count statistics.

Statistical noise in the images has been assessed by calculating the average relative standard deviation in a ROI located in the homogenous (i.e. lesion-free) part of the liver. The contrast was addressed using the $SUV_{max}$ values calculated according to eq. (24). Fig. 13 shows the *iRDF* convergence of the $SUV_{max}$ for the lung tumor and the relative standard deviation in the liver region.

## IV. DISCUSSION

### *A.* One-time iRDF calibration

Optimal detectability of small objects (here with 10 mm diameter) was found for $\alpha$ values between 2 and 3, as illustrated in the CRNC results shown in left column of Fig. 5. Increasing α beyond 3 led to slightly improved image contrast recovery and thus quantitatively more accurate results (see Fig. 5, right column).

Counter-intuitively, the reference contrasts measurements without *iRDF* displayed a slight improvement for smaller sphere radii in combination with reduced acquisition time. One potential reason for this might be a reduced accuracy of the NEMA characterization approach due to higher statistical fluctuations in the reconstructed voxel values.

While the low-count images indicate superior NEMA contrast values, the original location and shape of small spheres can hardly be determined from the reconstructed images. Therefore, CRC does not seem to be a reliable measure for benchmarking low-count data sets. Comparing non-regularized versus regularized results, feature detectability as defined in (20) and (21) was found to be about 3 times higher using *iRDF,* more or less independent from sphere size and α selection. The qualitative improvement can also be visually confirmed in Fig. 6, where we directly compare different image qualities achieved for a wide acquisition time range.

For $\alpha>4$, a number of clustered noise spots can be found in the images, resulting from (compared to the homogeneous background) wrongly characterized *features* (edges) in the *iRDF* noise prior term. Consequently, although contrast further improves with increasing α values, the spatial noise also increases, leading to a further reduction in feature detectability (compare left and right column in Fig. 5). In order to support a reliable identification of even smaller lesions via enhanced lesion detectability and feature contrast, we choose $\alpha=3$ as the standard setting for *iRDF.*

Furthermore, it can be appreciated from Fig. 6 (column 4) that when using *iRDF*, the relative image quality and noise characteristics remained at a similar level when mimicking an acquisition time reduction from 180s to 45s (i.e. $\sim 1.1 \cdot 10^7$ detected coincidences), especially in contrast to the standard non-regularized reconstruction outcome (column 1).

With higher α values, however, the likelihood for artificial hot-spots clearly increases (see also top rows in Fig. 7). The visual impression also supports the former hypothesis (section II.E) that the *RDP* approach intrinsically suffers from hot-spot artifacts especially in object areas with weak count statistics. Thus, noise clusters show a higher tendency to form towards the axial FOV boundaries due to diminished scanner sensitivity in these areas.

### *B.* Detection and correction of artificial hot-spots

The outcome illustrates a clearly positive effect regarding artificial feature suppression. However, even in the preferred $\alpha=3$ case not all artifacts have been removed. All remaining artifacts except a single one (highlighted in Fig. 8) are again located at the axial FOV boundary. This visual impression again confirms that (in high contrast configuration) the *RDP* approach is likely to produce noise-artifacts in regions with low count statistics. The remaining central hot-spot in the high sensitivity region was found to actually be formed by 2 blobs. As illustrated in Fig. 8, applying the second-max penalization approach in *iRDF* also effectively suppressed this remaining artifact.

## *C.* Minimum number of detected decays per image element needed for RDP

The analysis of the NEMA IQ phantom data reconstructed using *iRDF* ($\alpha=3$), indicated a minimum effective threshold of $\gamma^*_{min} = 13.5$, which translates to a lower boundary of

$$n_{min} = \left(\frac{\gamma^*_{min}}{\alpha}\right)^2 = (13.5/3)^2 = 20.25 \text{ counts per blob.} \quad (25)$$

Therefore, we deduce that in our scanner and reconstruction setup at least ~20 counts per blob are required in order to reliably apply *iRDF*.

## *D.* Evaluation using NEMA IQ phantom and clinical data

Results presented in Fig. 10 show a clear improvement regarding noise pattern suppression. Also, as spatial resolution appears unaffected, a more evident – and probably also earlier – lesion detectability can be expected. Contrast is found to be increased for higher acquisition times (e.g. 180s), since more iterations have been performed in *iRDF* (500 updates) than in the OSEM reconstruction (5x17=85 updates) without regularization prior. For very low data statistics (< 20 counts per blob) contrast is found to be diminished, as resulting lower effective gammas (see eq. (23)) led to an overall intensified spatial smoothing with quadratic prior component of *iRDF*. However, as the noise cancellation effect is clearly more dominant, an improved detectability as defined by eq. (20) and (21) can be observed even for (rather) short acquisition times.

Comparison between *iRDF*, standard *RDP*, and non-regularized OSEM in Fig. 9 illustrates that *RDP* parameter γ moderates resulting image quality between two modes: a) preserving contrast of OSEM, but only slightly reducing noise in homogenous regions (Fig. 9, column 2), and b) improving noise reduction but at the cost of reducing spatial resolution in small regions (Fig. 9, column 3). In contrast to this, results with *iRDF* show that the local automatic choice of penalty strength, once calibrated for α, leads to both high contrast and high noise reduction (Fig. 9, column 4).

These visual results are confirmed by quantitative measures shown in Table I for 180s acquisition time and Table II for 45s:

a) $SUV_{max}$ (*maximum standardized uptake value*) inside the small regions is confirmed to be high with low regularization (i.e. standard OSEM and *RDP* with high values for γ) and with *iRDF*.

b) Standard deviation in the background region shows opposite behavior in case of OSEM and *RDP*, but not for *iRDF*.

With respect to the convergence behavior of *iRDF*, Fig. 11 indicates that the initial convergence speed highly depends on the size of the corresponding feature, which results in a faster convergence of the $SUV_{max}$ in larger active regions (e.g. with 22 mm diameter) compared to smaller regions such as in the case of the 10 mm sphere. $SUV_{max}$ for spheres ≥13 mm diameter was very similar for both 180s and 45s acquisition time, while for the 10 mm sphere, the reduction of count statistics by a factor of four led to a $SUV_{max}$ reduction of ~30% (from 11.6 for 90s to 8.2 for 22.5s). In this context we would like to highlight that due to the noisy nature of these images, even when using *iRDF* there is a high variability in $SUV_{max}$ which can cause higher local values with 45s than with (more accurate) 180s measurements. Especially for non-regularized MLEM the high image noise leads to worse reproducibility and self-consistency of $SUV_{max}$ (compare 2$^{nd}$ column of Table I with Table II). Here, the relative standard deviation could be used as an indicator for the reliability of the reconstruction results.

As with the NEMA IQ phantom outcome, the clinical images in Fig. 12 also illustrate a significantly improved noise control, while the lesion contrast in terms of $SUV_{max}$ values has been effectively preserved with *iRDF*. In contrast, with *RDP*, the constant parameter γ cannot be set to a unique value that can provide both a good contrast and noise reduction for all regions of the FOV. Compared to standard OSEM, a slightly improved $SUV_{max}$ could be observed in the 90s acquisition due to the larger number of image updates (500 iterations of MLEM vs. 5x17 iterations of OSEM) which led to a better contrast recovery. The *iRDF* convergence shown in Fig. 13 seems to reach stable state after ~200 iterations as already observed for the NEMA phantom study in Fig. 11. For the 22.5s acquisition, $SUV_{max}$ was reduced by all regularization approaches despite the higher number of image updates.

The proposed method of *iRDF* has been benchmarked with MLEM as the most stable and "pure" form of iterative image reconstruction. For practical reasons (mainly computational time), we expect that for a more practically oriented *iRDF* implementation faster reconstruction methods such as BSREM or relaxed OSEM will be favored.

## V. CONCLUSIONS

We addressed in our investigation a common issue related to most current regularization approaches in PET image reconstruction, namely the cumbersome process of case-individual (and sometimes multi-parametrical) optimization. The proposed information-adaptive parametrization *iRDF* method inspired by standard *RDP* demonstrated its potential to generally overcome this burden. Our analyses also showed that with standard *RDP*, it was difficult to achieve both good contrast and noise mitigation at the same time. Compared to that, *iRDF* effectively adapts the tradeoff decision between high contrast and increased smoothness based on estimated local count statistics. One drawback of *iRDF* method is that by introducing the image estimate dependent edge preservation threshold parameter, it deviates from the classical MAP reconstruction theory. Nevertheless, *iRDF* demonstrated an overall stable practical convergence in terms of SUV recovery, allowing extension of the iterative data processing in the reconstruction to sufficiently approach practical SUV recovery without risking related noise amplification effects. In combination, this bears the potential to simplify/unify the clinical PET workflow, supporting also inter- and intra-departmental comparability in clinical studies.

The experiments and parameter analysis clearly confirm the general benefit of *RDP*-like (edge-preserving) noise regularization approaches for PET. However, they also illustrate the effect of the intrinsic limit below which the acquired spatial information density (represented by the number of detected decays per image element) does not allow reliable distinction between spatially clustered random noise and real, small (and weakly active) lesions. Statistical noise in the acquired list-mode data passing this regularization-parameter-specific boundary can lead to artificial features in the reconstructed image. These artifacts do not only visually degrade the image quality but (due to the enhanced background homogeneity) may even lead to increased likelihood of false positive readings. First results achieved with related *iRDF* modifications significantly improved hot-spot control with only minor lesion contrast impairment. Further optimization and validation of these methods will be addressed in follow-up investigations.

In general, *iRDF* provides an effective approach to improve overall image quality while preserving clinically relevant study parameters, such as SUV-max, which is frequently used in quantitative therapy response monitoring. Moreover, according to its stable performance under acquisition time/dose reduction, *iRDF* may be considered an interesting option especially for fast-scan/low-dose PET imaging applications, as well as for the purposes of harmonizing the quantification of PET studies.

Follow-up studies are suggested to investigate how to adequately apply *iRDF* concepts in an adapted ordered-subset scheme in order to efficiently accelerate image formation. Additionally, the hot-spot correction method needs to be revised for improving the convergence of small lesions e.g. by replacing the binary correction strategy used in eq. (22) by a smoother transition penalty to avoid ladder-like effects during convergence as observed in Fig. 11.

Moreover, fixed scanner sensitivity will have to be replaced by a TOF and object dependent effective sensitivity to reduce inter-system stability of parameter $\alpha$. Finally, *iRDF* will be tested on larger amount of clinical data sets to further investigate applicability in clinical environment.

## ACKNOWLEDGMENT

We would like to thank the group of M. V. Knopp, and J. Zhang from the Wright Center of Innovation in Biomedical Imaging, Department of Radiology, The Ohio State University Wexner Medical Center, Columbus, OH, as well as Piotr Maniawski from Advanced Molecular Imaging, Philips, Cleveland, OH for acquiring and providing the clinical data. Finally, we would like to thank Cloay Maier, also from Advanced Molecular Imaging, Philips, Cleveland, OH for her support in reviewing the manuscript.